\documentclass[letter,letterpaper]{aa}
\usepackage{txfonts}
\usepackage{graphicx}
\usepackage{natbib}
\usepackage[T1]{fontenc}
\bibpunct{(}{)}{;}{a}{}{,} %to follow the A&A style

\begin{document}

\author{Gu{\dh}laugur J{\'o}hannesson\and Gunnlaugur Bj{\"o}rnsson \and Einar H. Gudmundsson}
\title{Luminosity Functions of Gamma-Ray Burst Afterglows}

\institute{Science Institute, University
of Iceland, Dunhaga~3, IS--107 Reykjavik, Iceland}

\date{Received / Accepted}

	\abstract{}{Use the standard fireball model to create virtual populations of
	gamma-ray burst afterglows and study their luminosity functions.}{We
	randomly vary the parameters of the standard fireball model to create
	virtual populations of afterglows.  We use the luminosity of each burst at an
	observer's time of 1 day to create a luminosity function and compare
	our results with available observational data to assess the internal
	consistency of the standard fireball model.}{We
	show that the luminosity functions can be described by a function similar to a
	log normal distribution with an exponential cutoff.  The function
	parameters are frequency dependent but not very dependent on the model
	parameter distributions used to create the virtual populations.
	Comparison with 
	observations shows that while there is good general agreement with the
	data, it is difficult to explain simultaneously the X-ray and
	optical data.  Possible reasons for this are discussed and the
	most likely one is that the standard
	fireball model is incomplete and that decoupling of the X-ray and optical
	emission mechanism may be needed.}{}


\maketitle

\section{Introduction}
Since its discovery in 1997
\citep{Costa1997,Paradijs1997}, the standard
fireball model, or extensions thereof, has been used to model gamma ray burst (GRB)
afterglow emission \citep[e.g.][]{Panaitescu2005, Johannesson2006a}.
With dedicated follow-up observations, the number of detected
afterglows has been steadily increasing.
The rapid burst localisation of the {\em Swift} satellite and its on-board
X-ray (XRT) telescope have substantially increased the afterglow
detection rate and the number of detected afterglows is now well over
200\footnote{see {\tt http://www.mpe.mpg.de/$\sim$jcg/grbgen.html} for
the most recent list}.  
%Even though observations with the XRT show
%that the early afterglow is more complex than previously thought, the
%standard fireball model is still able to explain the general afterglow behaviour
%several hours post-burst \citep{Nousek2006}.
This
has inspired statistical investigations of
afterglow properties; among which are studies exploring the
luminosity function at a given observer time
\citep{Gendre2005,Kann2006,Liang2006,Nardini2006,Nardini2007}.

Motivated by this previous work, we have created theoretical
luminosity functions of GRB afterglows using the standard fireball
model.  We show that these
luminosity functions can be described to a high degree of accuracy by an
analytical function.  Furthermore, we show that the parameters of that
function are
for the most parts independent of the chosen model 
parameter distributions, giving a robust estimate of the luminosity
function of GRB afterglows.  We also show that our theoretical luminosity
function is similar to the observationally determined one, both in
X-rays and optical wavebands but only if they are considered separately.  It is difficult to
explain the observed luminosity functions simultaneously in both wavebands.
Several possible reasons
for this are discussed, the most likely one being that the standard
fireball model is incomplete.

The outline of this letter is as follows: The procedure for creating the
luminosity functions is described in section 2 and we present our
results and discuss how they depend on the model parameters in section
3.  Section 4 is devoted to comparison with observations and we conclude
the letter in section 5.

\section{A Virtual World of Afterglows}
The afterglows are created numerically with the model described in
\citet{Johannesson2006a}.  It is based on the standard
fireball jet model, in which the energy is injected instantaneously into
a narrow jet.  
To create the afterglows, model parameters are selected at random from
pre-defined distributions \citep[see also][]{Johannesson2006b}.  Each
parameter range is determined so that
it represents results found by fitting afterglow observations with the
standard fireball model or extensions thereof \citep[e.g.][]{Panaitescu2001, Panaitescu2005,
Johannesson2006a}.  Table \ref{tab:parameter_ranges} shows the
default parameter ranges used in this work.  The parameter values can in many
cases vary by more than an order of magnitude from one afterglow to
another.  All model
parameters, except $p$, are therefore varied on a logarithmic scale to get a more
realistic parameter distribution within the chosen range. To concentrate on the intrinsic properties of
the afterglows, we fix the redshift at $z=1$.

\begin{table}
\caption{Default values for the upper and lower limit of the parameter
	ranges.  All parameters except for $p$ are distributed
	logarithmically in the interval.}
\label{tab:parameter_ranges}
	\centering
\begin{tabular}{l c c}
	\hline
	\hline
	Parameter & Lower limit & Upper limit \\
	\hline
	$E_0$ [erg]         & $10^{49}$ & $10^{51}$ \\
	$n_0$ [cm$^{-3}$]   & 0.1       & 10 \\
	$\theta_0$ [$\deg$] & 2         & 15 \\
	$p$                 & 2.1       & 2.6 \\
	$\epsilon_e$        & 0.05      & 0.3 \\
	$\epsilon_B$        & $10^{-5}$ & $10^{-3}$ \\
	\hline
\end{tabular}
\end{table}

To calculate the luminosity functions, we use the afterglow
luminosity density at an observer's time of 1 day post-burst.  In fact, any time can be used, but
this choice is well suited for comparison with observations
\citep{Liang2006, Nardini2006}.  One day after the burst is also late
enough in the afterglow evolution so that both the optical and X-ray
emission can be assumed to originate in
the forward shock of the standard fireball model \citep{Nousek2006}.  To reduce
statistical errors, each virtual world consists of 50,000 afterglows,
resulting in statistical errors of the order of 1\% in the calculated
luminosity functions.

\section{Luminosity Functions}

\begin{figure}
	\resizebox{\hsize}{!}{\includegraphics{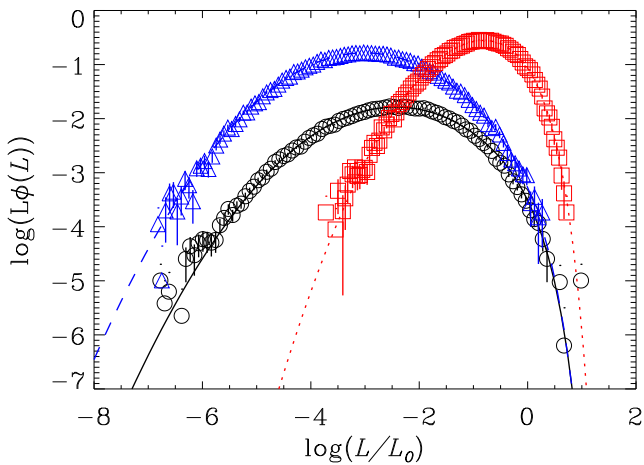}}
\caption{Example luminosity functions obtained in our numerical
	calculations at three different frequencies: 2-10 keV (circles, solid
	curve),
	$R$-band (triangles, dashed curve, blue in electronic edition) and 8.5
	GHz (squares, dotted curve, red in electronic edition).  The curves are best fits using
	equation~(\ref{eq:phi}).  For clarity we plot $L\phi(L)\times0.1$
	for 2-10 keV.  Note that the general shape of the
	luminosity function is frequency independent although $\lambda$,
	$\sigma$ and $L_0$ are not.}
\label{fig:pdf}
\end{figure}

\begin{figure}
\resizebox{\hsize}{!}{\includegraphics{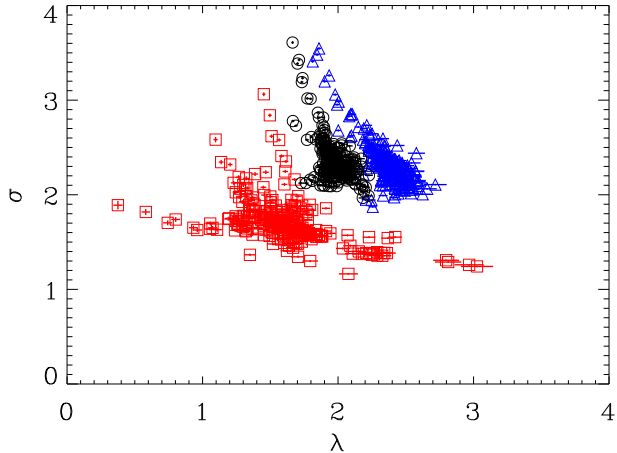}}
\caption{The plot shows that there is an anti-correlation between the
	values of $\lambda$ and $\sigma$ found from fitting the numerically
	calculated luminosity functions with equation~(\ref{eq:phi}).  The
	Pearson's correlation coefficient varies from -0.5 to -0.8 depending
	on frequency, being strongest in the optical range and weakest in
	radio.  See fig.~\ref{fig:pdf} for legend.}
	\label{fig:sigma_lambda_correlation}
\end{figure}

Figure \ref{fig:pdf} shows typical luminosity functions at three
different observer frequencies.  Overlaid on the numerical calculations
are fits to the results using the function
\begin{equation}
	\label{eq:phi}
	\phi(L) = C\left( \frac{L}{L_0}
	\right)^{-\lambda}\exp{\left(-\frac{\ln\left( L/L_0
	\right)^2}{2\sigma^2}
	\right)}\exp\left({-\frac{L}{L_0}}\right).
\end{equation}
We have chosen this function because of the parabolic resemblance of the
plot shown in fig.~\ref{fig:pdf}.  The last exponential factor is needed
since the luminosity function is not a pure parabola in log-log space.
Here, $L_0$ is a characteristic luminosity, defining an upper
exponential cutoff in the function.  Both $\sigma$ and $\lambda$
affect the width of the function, controlling its shape, while
$\sigma$ has a slightly stronger effect.  $C$ is a normalization
constant and is determined by the other 3 parameters.
Our results show that $\phi(L)$ is an accurate description of
the numerical luminosity functions, giving a $\chi^2$ between 50 and
200 in each case for 100 degrees of freedom.  Note that the function becomes a
log-normal distribution with an exponential cutoff when $\lambda=1$. 
Our results show, however,  that a value of $\lambda \approx 2$ is in the present
case more likely,
varying between the values 1.5 and 2.5, depending on frequency and input model
parameter
distributions (see fig.~\ref{fig:sigma_lambda_correlation}).

To investigate how the model parameter distributions affect our results, three different
distribution shapes were tested, a uniform one, a Gaussian one and a
triangular
distribution with a maximum in the center of the parameter range.  In
the case of the Gaussian, the standard deviation was set at 1/4th
the parameter range and the distribution restricted to the defined
range.  We
found that the shape of the model parameter distributions did not change our
general results, although the exact values of the parameters in
equation~(\ref{eq:phi}) do depend on it.  Since the function parameter
values change only slightly between the chosen input distributions, we
decided to use the
triangular distribution for all our results below.

\begin{figure*}
	\centering
\includegraphics[width=17cm]{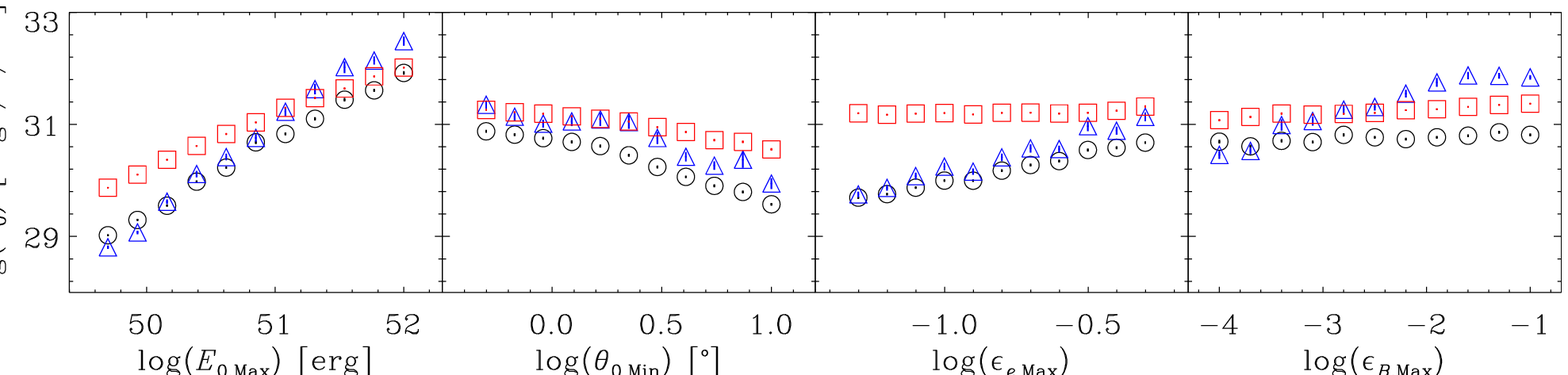}
\includegraphics[width=17cm]{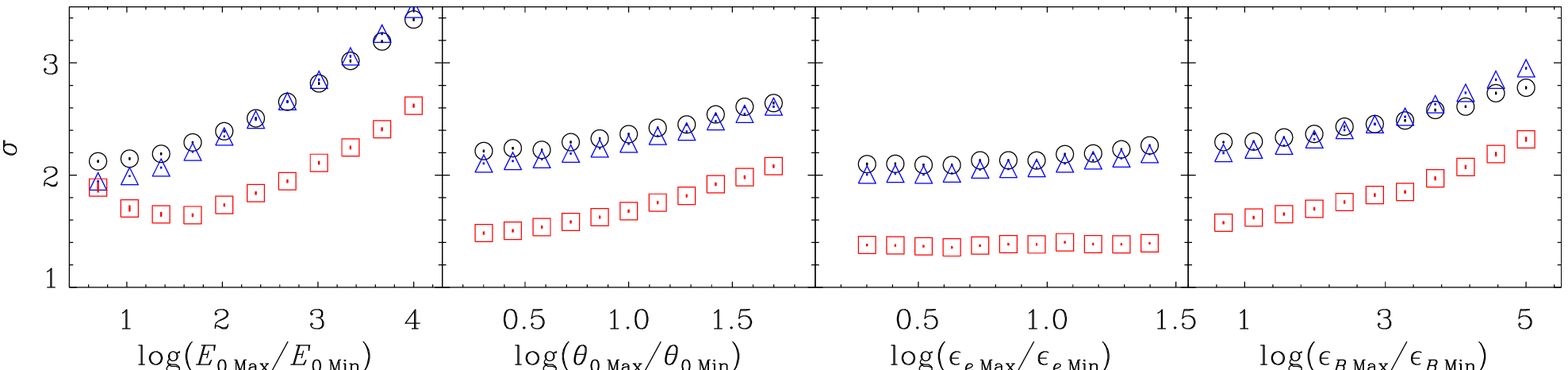}
\caption{Shown are the values of some of the function parameters
	found from fitting numerically calculated luminosity functions with
	equation~(\ref{eq:phi}).  See fig.~\ref{fig:pdf} for legend.
	In each case we vary the limits of the model
	parameter distribution and show only the parameters that have the
	strongest effects.  The default parameter limits are given in
	table~\ref{tab:parameter_ranges}.
	{\em Top panel:}  The value of $L_0$ while varying the
	upper limit (lower limit in the case of $\theta_0$) of the model parameter distribution keeping the lower
	limit (upper limit in the case of $\theta_0$) fixed.  Shown are results for the parameters $E_0$, $\theta_0$,
	$\epsilon_e$ and $\epsilon_B$.  We multiply $L_0$ by $10^3$ for 2-10
	keV.  
	{\em Bottom panel:} The value of $\sigma$ while varying the width of
	the model parameter distribution, keeping the center fixed.  Shown are
	results for $E_0$, $\theta_0$, $\epsilon_e$ and
	$\epsilon_B$.  Since the parameters are
	changed logarithmically, the width is the ratio between the upper and
	lower limit. 
	}
\label{fig:variations}
\end{figure*}

In addition to varying the shape of the model parameter distributions, we also
tested how the parameter limits affect the outcomes.  In each case, we
changed the limits of only one parameter at a time.
To cover all the basic effects, four tests were made: first we changed the upper limit keeping the
lower limit fixed, next the lower limit was varied keeping the upper
limit fixed, then the
width was changed with a fixed center and finally the center was changed with a fixed width.  
The default values for the upper and lower limits are shown in
table~\ref{tab:parameter_ranges}.  All of the parameters, except $p$, are
distributed logarithmically and the width is then defined as
the ratio between the upper and lower limit.
Figure~\ref{fig:variations} shows the most interesting results from our
tests.  The parameters found from fitting
the numerically calculated luminosity function with
equation~(\ref{eq:phi}) are plotted against the parameter that is being
varied in each case.  The general result from these tests, is that the function parameters
$\lambda$ and $\sigma$ are not very dependent on the limits of the model
parameter distributions, while $L_0$ can vary by  several orders of magnitude.  This allows
us to conclude that the luminosity function can be presented by
equation~(\ref{eq:phi}), setting both $\lambda\sim2$ and $\sigma\sim2$ and
using $L_0$ as the only variable.

We see from fig.~\ref{fig:variations} that
the most significant model parameter is $E_0$, the initial energy
release.  Other
model parameters that have significant effects on the
parameters of $\phi(L)$ are $\theta_0$, the initial opening angle of the jet, and
$\epsilon_e$ and $\epsilon_B$, the fractions of energy contained in the
electron population and the magnetic field, respectively.  The effects of
the external density $n_0$ and the electron energy distribution index,
$p$, on the function parameters are insignificant compared to the other
model parameters.

The general trend when varying the parameter limits is
that the value of $L_0$ is correlated with the upper limit (lower limit
in case of $\theta_0$) of the
model parameter range, while $\sigma$ is correlated with the width of the
parameter distributions.  This is shown, respectively, in the top and
middle panel of 
fig.~\ref{fig:variations}.  Note that the changes in the
parameters of $\phi(L)$ depend in many cases strongly on the observed frequency. This
can, for example, be seen in the value of $L_0$ while varying the upper limit of the $\epsilon_e$.  There are no
effects in the radio waveband whereas the effects in optical and X-ray
wavebands are similar and quite significant.  Although the value of $\sigma$
depends most strongly on the width of the
model parameter distribution, it can in some cases also
depend on the center value of the model parameter distribution. 
%as is shown in the bottom panel in figure~\ref{fig:variations}.  
The trend
between the center value and $\sigma$ is, however, not as strong as the
trend between $\sigma$ and the width.  

The function parameter that shows the least dispersion when varying the
limits of the model parameter distributions is $\lambda$.  Its value is generally
clustered around 2 and shows a greater dependency on frequency than
on the model parameters, while its frequency dependence is in fact similar
to that of $\sigma$ and $L_0$.  Although $\lambda$ shows the least
dispersion when varying the model parameter ranges, it is the parameter that is
most difficult to constrain in the fitting procedure, resulting in the
highest 1-sigma error.
This can be seen in fig.~\ref{fig:sigma_lambda_correlation}, which shows the value of $\sigma$ as
a function of $\lambda$.  It is also clear from the figure that there is an
anti-correlation between the values of $\sigma$ and $\lambda$.
Pearson's correlation test confirms this and results in correlation
parameters between -0.5 to -0.8 depending on frequency.  The correlation
is strongest in the optical but weakest in the radio.

\section{Comparison with Observations}

\begin{figure}
\resizebox{\hsize}{!}{\includegraphics{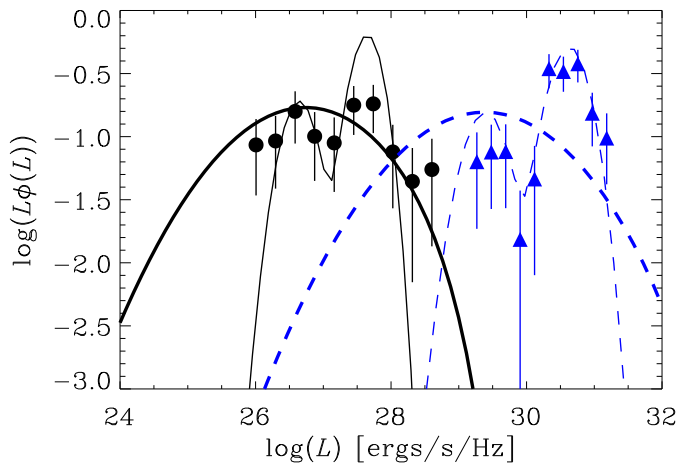}}
\caption{Comparison between our numerical calculations (curves) and
	luminosity functions obtained from observational data
	\citep{Nardini2006} (filled symbols).  Two different models are shown,
	one wide (thick curves), assuming the value of the center of the $E_0$ distribution is at
	$10^{51}$ erg and its width at the default value of 100.  Other
	parameters are as in 
	table~\ref{tab:parameter_ranges}.  The other model is a composite luminosity
	function of two
	different populations, each with a much narrower parameter distribution
	(thin curves).  Each population has, in this case, a different value for the center
	of the $E_0$ distribution and a different number of afterglows but
	are otherwise the same.  See text for further discussion.  Two different frequencies: 2-10
	keV (circles, solid curves) and $R$-band (triangles, dashed curves,
	blue in electronic edition) are shown.}
	\label{fig:observations}
\end{figure}

To compare our results with observations we took the recent luminosity
compilation from \citet{Nardini2006} and converted it to luminosity
functions using the same method as for our numerical results.  This data is suitable for our use, since it
is normalized to an observer's time of 1 day and a redshift of 1.  The data
is displayed in
fig.~\ref{fig:observations} along with selected luminosity
functions from our calculations.  Note that our results are not formal
fits to the observational data, but rather an eye-guided selection of
one of our calculated luminosity functions.  The sparse data set does not
warrant a formal fit.  Selection effects are also not taken into account for the same
reason, but we refer interested readers to \citet{Nardini2007} which
discusses the selection effects.
The model parameter distribution that best resembles the observed X-ray luminosity function
is the one with the center of the $E_0$ distribution at $10^{51}$ erg
but its width at the default value of 100.  Other
parameters were at their default values, given in
table~\ref{tab:parameter_ranges}.  This energy distribution is
comparable to the one found from observations by \citet{Frail2001}.
The resulting luminosity function is, however, incompatible with the observed
optical luminosity function that shows a clear over abundance of high
luminosity afterglows which can not be explained by our results.  There
is also a sign of bimodality in the observed optical luminosity function \citep{Liang2006,
Nardini2006, Kann2006, Jakobsson2006}.  

In an attempt to understand the optical luminosity function, we also
created two afterglow populations with very narrow model parameter
distributions, the width being a factor of 2 to 3.  These distributions are
not compatible with parameter ranges found from fitting afterglow observations
with the standard fireball model, where the ranges seem to be much
larger.  
The two populations differ only in the value of the center of
the $E_0$ distribution and the number of afterglows in each population.
The more luminous population has the center value of $E_0$ at $10^{51}$
ergs
while the fainter has the center at $2\cdot10^{50}$ ergs.  The number of afterglows in the
more luminous population is four times higher than in the fainter one. 
Note that this ratio may change if observational biases are taken into 
account, as discussed below.  These
populations give two very narrow luminosity functions, which when added
together agree with the observed
optical luminosity function.  It does not, however, seem to agree with
the X-ray data, although a slight hint of a high luminosity peak may be
seen.  Its statistical significance is unknown at present.

There are a few possible explanations for the difference between the
optical and X-ray luminosity functions.  First, there may be two populations of afterglows with
different parameter distributions, as shown by the thin
lines in fig.~\ref{fig:observations}.  Since this gives a good
representation of the bimodal optical luminosity function and there is a
hint of  bimodality in the
observed X-ray luminosity function, as discussed above, this explanation
may seem to agree better with the observational data than a single
population would.  However, the narrow luminosity functions
required for this explanation force the model parameter
distributions to be much narrower than those found
from afterglow modeling.  Tight constraints between model parameters might, however, result in a
similarly narrow luminosity functions and still comply with afterglow
modeling.  But even if that is the case, the ratio between the
two peaks in each observed luminosity function is different and a fit
with two afterglow populations would therefore be
marginally good at best.  

The second possibility is that the standard
fireball model needs to be refined.  Since there is a distinct
difference between the optical and X-ray luminosity functions, a
decoupling of the X-ray and optical radiation mechanism may be needed.  One
example could be that the X-ray and optical radiation
originate in different parts of the outflow.  The bimodality of the
optical luminosity function might then originate in two different
populations of afterglows.  The lack of bimodality in
the X-ray luminosity function would then be accounted for by a larger
dispersion in the radiation power of the X-ray zone. This possibility is also
supported by recent observation of GRB afterglows \citep[e.g.][]{Perley2007,
Oates2007} that show incompatibility with 
the standard fireball model because of differences between the X-ray and 
optical light curves.

Finally, the observed optical
luminosity function may be incomplete.
Limitations of current optical follow-up instruments, either in 
sensitivity or response time, lead to an observational bias for luminous
afterglows.  The bimodality is then expected to vanish
with increasing number of observations.  This may
already have happened with the X-ray afterglow
observations, as \citet{Gendre2005} found early on an evidence for bimodality in
the X-ray luminosity function, which can now barely be seen with the more
rapid and sensitive observations.  If not real, the bimodality in the
optical luminosity function may require a detection of a larger number of afterglows
before it disappears, because of the redshift dependence of the data.  
The redshift is in most cases acquired from the optical afterglow
spectrum and therefore requires an optically bright afterglow.  This further
increases the
bias for optically bright afterglows without affecting the X-rays much.
The recent study by \citet{Nardini2007} has shown that this is an unlikely
explanation.  The luminosity functions are, however, still based on a very small sample.
In addition, there are also uncertainties involved when doing the K-correction and 
time shift of the observational data.
Hence, we cannot completely rule this explanation out.

\section{Discussion and Conclusions}

In this letter we have shown, with the help of numerical calculations, that the
luminosity functions of GRB afterglows can be described by a rather simple
function given by equation~(\ref{eq:phi}).  It is similar to a log normal distribution with an
exponential cutoff.  We have also shown
that the shape of the luminosity function is not sensitive to the
shape of the model parameter distributions.  The function parameters are
also quite robust, $\lambda$ and $\sigma$
vary by less than a factor of 2, in most cases only by a few percent,
for variations of orders of magnitude in the model
parameters.  It is mainly the upper cutoff, $L_0$, that is affected
by the input parameters.  This result can be used to get an approximate single
parameter representation of the afterglow luminosity function.  One
must, however, bear in mind that
$\sigma$ and $\lambda$ are somewhat dependent on our
assumptions for
the model parameter distributions.  While we have chosen those to be representative
of parameter estimates from current afterglow modeling, future 
modeling may require a re-evaluation of the values of $\lambda$
and $\sigma$.  However, these are not expected to change much unless
significant changes are seen in the model parameter
distributions or strong correlations are found between some of the model parameters.

We have also shown that our results are in good general agreement with
observations.  However, incompatibility apparently exists between the optical and
X-ray data, where a bimodality is seen in the optical luminosity
function that is not apparent in the X-rays.  The most likely
explanation for this is that the standard fireball model is an
incomplete description of the afterglow emission.  A
more detailed model or
possibly an entirely different one is required, where a
decoupling of the sources of X-ray and optical radiation is a natural ingredient.
This is supported by recent observations, which have shown the afterglow behaviour to be more
complex than expected \citep[e.g.][]{Nousek2006}.  

\begin{acknowledgements}
This work was partially supported by The Icelandic Research Fund and
the University of Iceland Research Fund.  We thank S. Courty for helpful discussions and P.
Jakobsson for critical comments on an early version of the manuscript.
\end{acknowledgements}

\end{document}